\documentclass{aa}
\usepackage{savesym}
\usepackage{amsmath}
\savesymbol{iint}
\usepackage{txfonts}
\restoresymbol{TXF}{iint}
\usepackage{graphics,graphicx}
\usepackage{amssymb}
\usepackage{color}
\usepackage[draft]{hyperref}
\usepackage{breakurl}
\usepackage{array}
\usepackage{rotating}
\usepackage{xcolor}
\usepackage{natbib}
\usepackage{float}
\usepackage{flafter}
\usepackage{booktabs}
\usepackage{afterpage}
\usepackage{lipsum}
\usepackage{longtable,pdflscape}
\bibpunct{(}{)}{;}{a}{}{,}

\newcommand{\SNR}{{MCSNR~J0512$-$6707}}

\newcommand{\chandra}{{\it Chandra}}
\newcommand{\xmm}{{\it XMM-Newton}}
\newcommand{\rosat}{{\it ROSAT}}

\newcommand{\spitzer}{{\it Spitzer}}

\newcommand{\vpshock}{{\tt vpshock}}

\newcommand{\vphabs}{{\tt vphabs}}
\newcommand{\vapec}{{\tt vapec}}
\newcommand{\phabs}{{\tt phabs}}
\newcommand{\pow}{{\tt powerlaw}}

\newcommand{\ha}{H$\alpha$}
\newcommand{\sii}{[$\ion{S}{ii}$]}
\newcommand{\oiii}{[$\ion{O}{iii}$]}
\newcommand{\ratio}{[$\ion{S}{ii}$]/H$\alpha$}

\begin{document}

\title{Multi-frequency study of the newly confirmed supernova remnant \SNR\ in the Large Magellanic Cloud\thanks{Based on observations obtained with \xmm, an ESA science mission with instruments and contributions directly funded by ESA Member States and NASA}}



\author{P.~J.~Kavanagh \inst{1} \and M.~Sasaki \inst{1}  \and L.~M.~Bozzetto \inst{2} \and S.~D.~Points\inst{3} \and M.~D.~Filipovi\'c \inst{2} \and P.~Maggi \inst{4} \and \\ F.~Haberl \inst{5} \and E.~J.~Crawford \inst{2}}

 

\institute{Institut f\"{u}r Astronomie und Astrophysik, Kepler Center for Astro and Particle Physics, Eberhard Karls Universit\"{a}t T\"{u}bingen, Sand~1, T\"{u}bingen D-72076, Germany\\ \email{kavanagh@astro.uni-tuebingen.de}
\and University of Western Sydney, Locked Bag 1791, Penrith, NSW 2751, Australia
\and Cerro Tololo Inter-American Observatory, Casilla 603, La Serena, Chile
\and Laboratoire AIM, CEA-IRFU/CNRS/Universit\'e Paris Diderot, Service d'Astrophysique, CEA Saclay, F-91191 Gif sur Yvette Cedex, France
\and Max-Planck-Institut f\"{u}r extraterrestrische Physik, Giessenbachstra\ss e, D-85748 Garching, Germany
}

\date{Received ?? / Accepted ??}

\abstract{}{We present a multi-frequency study of the supernova remnant \SNR\ in the Large Magellanic Cloud.}{We used new data from \xmm\ to characterise the X-ray emission and data from the Australian Telescope Compact Array, the Magellanic Cloud Emission Line Survey, and \spitzer\ to gain a picture of the environment into which the remnant is expanding. We performed a morphological study, determined radio polarisation and magnetic field orientation, and performed an X-ray spectral analysis.}{We estimated the remnant's size to be $24.9~(\pm1.5)\times21.9~(\pm1.5)$~pc, with the major axis rotated $\sim29^{\circ}$ east of north. Radio polarisation images at 3~cm and 6~cm indicate a higher degree of polarisation in the northwest and southeast tangentially oriented to the SNR shock front, indicative of an SNR compressing the magnetic field threading the interstellar medium. The X-ray spectrum is unusual as it requires a soft ($\sim0.2$~keV) collisional ionisation equilibrium thermal plasma of interstellar medium abundance, in addition to a harder component. Using our fit results and the Sedov dynamical model, we showed that the thermal emission is not consistent with a Sedov remnant. We suggested that the thermal X-rays can be explained by \SNR\ having initially evolved into a wind-blown cavity and is now interacting with the surrounding dense shell. The origin of the hard component remains unclear. We could not determine the supernova type from the X-ray spectrum. Indirect evidence for the type is found in the study of the local stellar population and star formation history in the literature, which suggests a core-collapse origin.}{\SNR\ likely resulted from the core-collapse of high mass progenitor which carved a low density cavity into its surrounding medium, with the soft X-rays resulting from the impact of the blast wave with the surrounding shell. The unusual hard X-ray component requires deeper and higher spatial resolution radio and X-ray observations to confirm its origin.}

\titlerunning{\xmm\ study of \SNR}
\maketitle 

\section{Introduction}
Supernova remnants (SNRs) are formed via the interaction of ejecta from violent supernova (SN) explosions with the surrounding medium. The supernovae (SNe) can result from either the core-collapse (CC) of a massive star or from the explosion of a carbon-oxygen white dwarf in a binary system which exceeds the Chandrasekhar limit, either via accretion from a stellar donor or white dwarf mergers, so-called Type~Ia explosions. The SN ejecta blasted into the interstellar medium (ISM) contain heavy elements produced in the stellar progenitors and via explosive nucleosynthesis which chemically enriches the ISM. The energy deposited by the explosions drive the mechanical evolution of the ISM and cosmic rays can be accelerated in the fast moving shocks \citep[see][for a review]{Vink2012}. Studies of these objects in the Milky Way are problematic due to uncertain distance estimates and high foreground absorption, which particularly affects evolved SNRs whose low temperature plasmas are extremely susceptible to foreground absorption. 

\par The \object{Large Magellanic Cloud} (\object{LMC}) is an excellent galaxy for the study of SNRs due to its ideal observational properties. At a distance of 50~kpc \citep{diBen2008} it is sufficiently close that its stellar population and diffuse structure is resolved in most wavelength regimes. The LMC is almost face-on \citep[inclination angle of 30--40$^\circ$,][]{vanderMar2001,Nikolaev2004} and the modest extinction in the line of sight (average Galactic foreground $N_{\rm{H}} \approx 7 \times 10^{20}$~cm$^{-2}$) means optical and X-ray observations of SNRs are only slightly affected by foreground absorption, whereas its location in one of the coldest parts of the radio sky \citep{Haynes1991} allows for improved radio observations without interference from Galactic emission. The total number of confirmed SNRs in the LMC now stands at  $\sim60$ \citep[][and references therein]{Maggi2015,Bozzetto2015}. Recent additions include \citet{Bozzetto2014}, \citet{Maggi2014}, \citet{Kavanagh2015}, \citet{Kavanagh2015b}.

Typically, objects are classified as SNRs based on satisfying certain observational criteria. The Magellanic Cloud Supernova Remnant (MCSNR) Database\footnote{\burl{http://www.mcsnr.org/}} requires that at least two of the following three observational criteria must be met: significant \ha, \sii, and/or \oiii\ line emission with an \ratio\ flux ratio $>~0.4$ \citep{Mathewson1973,Fesen1985}; extended non-thermal radio emission; and extended thermal X-ray emission. A discussion of the significance of each of these classification criteria is given in \citet{Filipovic1998}. 

The \rosat\ catalogue of X-ray sources in the LMC of \citet{Haberl1999} contained the source \object{[HP99]~483} at a J2000 position of RA~=~05$^{\rm{h}}$12$^{\rm{m}}$28.0$^{\rm{s}}$ and Dec~=~$-67$$^{\rm{d}}$07$^{\rm{m}}$27$^{\rm{s}}$ which is located in the wider DEM~L97 HII region \citep{Davies1976}. The source is also projected near the edge of the molecular cloud \object{[WHO2011]~A126} \citep{Wong2011} to its south and southwest. However, due to the faintness of the source, only coarse properties of [HP99]~483 were determined by \citet{Haberl1999} and the source did not make it into the final list of \rosat\ detected SNRs and candidate SNRs. 

The study of \citet{Reid2015} identified the optical counterpart to [HP99]~483 using deep \ha\ imaging of the LMC performed with the United Kingdom Schmidt Telescope (UKST). Spectroscopic follow-up of several regions  in the object with the Anglo Australian Telescope and the 1.9~m telescope at the South African Astronomical Observatory revealed the characteristically high \ratio\ values, consistent with an SNR. The radio counterpart was also detected using data from the Australia Telescope Compact Array (ATCA), whose spectral index of $-0.52~(\pm0.04)$ is consistent with with the typical value of $-0.5$ for non-thermal emission from SNRs. Therefore, the object was confirmed as an SNR by \citet{Reid2015}, who assigned the identifier \SNR. A study of the surrounding dust emission using \spitzer\ data was also presented, with the authors noting that the SNR is located in a complex dust environment with the SNR appearing to be encased in dust, notably by a bright filament in the east and north. Using optical imaging, these authors determined the size of the remnant to be $13.5\times15.5$~pc and estimated the age to be 2--5~kyr assuming that the remnant is in the Sedov phase.

In this paper we present new X-ray observations of \SNR\ with \xmm. As we will show, the new X-ray data revealed that the size of \SNR\ is significantly larger ($24.9~(\pm1.5)\times21.9~(\pm1.5)$~pc, see Section~\ref{mwm}) than determined by \citet{Reid2015}, with the optical emission only present in the northeast of the remnant. In addition, we will show that \SNR\ is likely not in the Sedov phase (Section~\ref{xre}). We performed a similar study to assess the morphology and physical characteristics in light of these new findings, making use of the data from the Magellanic Cloud Emission Line Survey \citep[MCELS,][]{Smith2006}, radio data from ATCA, and IR data from \spitzer\ reported by \citet{Reid2015}.

Our work on the multi-frequency study of \SNR\ using X-ray, radio, IR, and optical emission line data is arranged as follows: the observations and data reduction are described in Section~2;  data analysis is outlined in Section~3; results are given and discussed in Section~4; and finally we summarise our work in Section~5.

\section{Observations and data reduction}
\subsection{X-ray}
\textit{XMM-Newton} \citep{Jansen2001} observed \SNR\ on May~17~2014 (Obs. ID 0741800201, PI P. Kavanagh). The primary instrument for the observation was the European Photon Imaging Camera (EPIC), which consists of a pn CCD \citep{Struder2001} and two MOS CCD \citep{Turner2001} imaging spectrometers. However, a ``false radiation'' warning due to a flaring optical source in the FOV affected the EPIC-MOS CCDs during this observation, resulting in both EPIC-MOS cameras being shut for extended periods. Because of this, the observation was repeated and a second dataset was obtained on June~8~2014 (Obs. ID 0741800301, PI P. Kavanagh). We reduced all available data (i.e., the usable data from Obs. ID 0741800201 and Obs. ID 0741800301) using the standard reduction tasks of SAS\footnote{Science Analysis Software, see \burl{http://xmm.esac.esa.int/sas/}} version 14.0.0, filtering for periods of high particle background. This resulted in combined flare-filtered exposure times of $\sim42$~ks for EPIC-pn, $\sim48$~ks for EPIC-MOS1, and $\sim46$~ks for EPIC-MOS2.

\subsection{Optical}
The MCELS observations \citep{Smith2006} were taken with the 0.6 m University of Michigan/Cerro Tololo Inter-American Observatory (CTIO) Curtis Schmidt Telescope equipped with a SITE 2048 $\times$ 2048 CCD, producing individual images of $1.35^{\circ} \times 1.35^{\circ}$ at a scale of 2.3$\arcsec$ pixel$^{-1}$. The survey mapped both the LMC ($8^{\circ} \times 8^{\circ}$) and the SMC ($3.5^{\circ} \times 4.5^{\circ}$) in narrow bands covering [\ion{O}{iii}]$\lambda$5007 \AA, H$\alpha$, and [\ion{S}{ii}]$\lambda$6716, 6731 \AA, in addition to matched green and red continuum bands. The survey data were flux calibrated and combined to produce mosaicked images. We extracted cutouts centred on \SNR\ from the MCELS mosaics. We subtracted the continuum images from the corresponding emission line images, thereby removing the stellar continuum and revealing the full extent of the faint diffuse emission. Finally, we divided the continuum subtracted \sii\ image by the continuum subtracted \ha\ image to get a \ratio\ map of \SNR, with regions of \ratio~$>0.4$ indicative of the presence of an SNR \citep{Mathewson1973,Fesen1985}.

\subsection{Radio}
\subsubsection{Radio continuum} 
\SNR\ was observed with the Australian Telescope Compact Array (ATCA) on November~15 and 16~2011 (project C634), using the new Compact Array Broadband Backend (CABB). The ATCA array configuration EW367 was used and observations were taken simultaneously at $\lambda = 3$ and 6~cm ($\nu$=9\,000 and 5\,500~MHz) using the dual-frequency mode. Baselines formed with the $6^\mathrm{th}$ ATCA antenna were excluded as the other five antennas were arranged in a compact configuration. The observations were carried out in the so called ``snap-shot'' mode, totalling $\sim$50 minutes of integration over a 14 hour period. PKS~B1934-638 was used for flux density calibration and PKS~B0530-727 was used for secondary (phase) calibration. The phase calibrator was observed twice every hour for a total 78 minutes over the whole observing session. The \textsc{miriad}\footnote{\burl{http://www.atnf.csiro.au/computing/software/miriad/}}  \citep{1995ASPC...77..433S} and \textsc{karma}\footnote{\burl{http://www.atnf.csiro.au/computing/software/karma/}} \citep{1995ASPC...77..144G} software packages were used for reduction and analysis.

\begin{figure}[!t]
\resizebox{\hsize}{!}{\includegraphics[trim=0cm 3cm 0cm 3cm,angle=-90,width=4in]{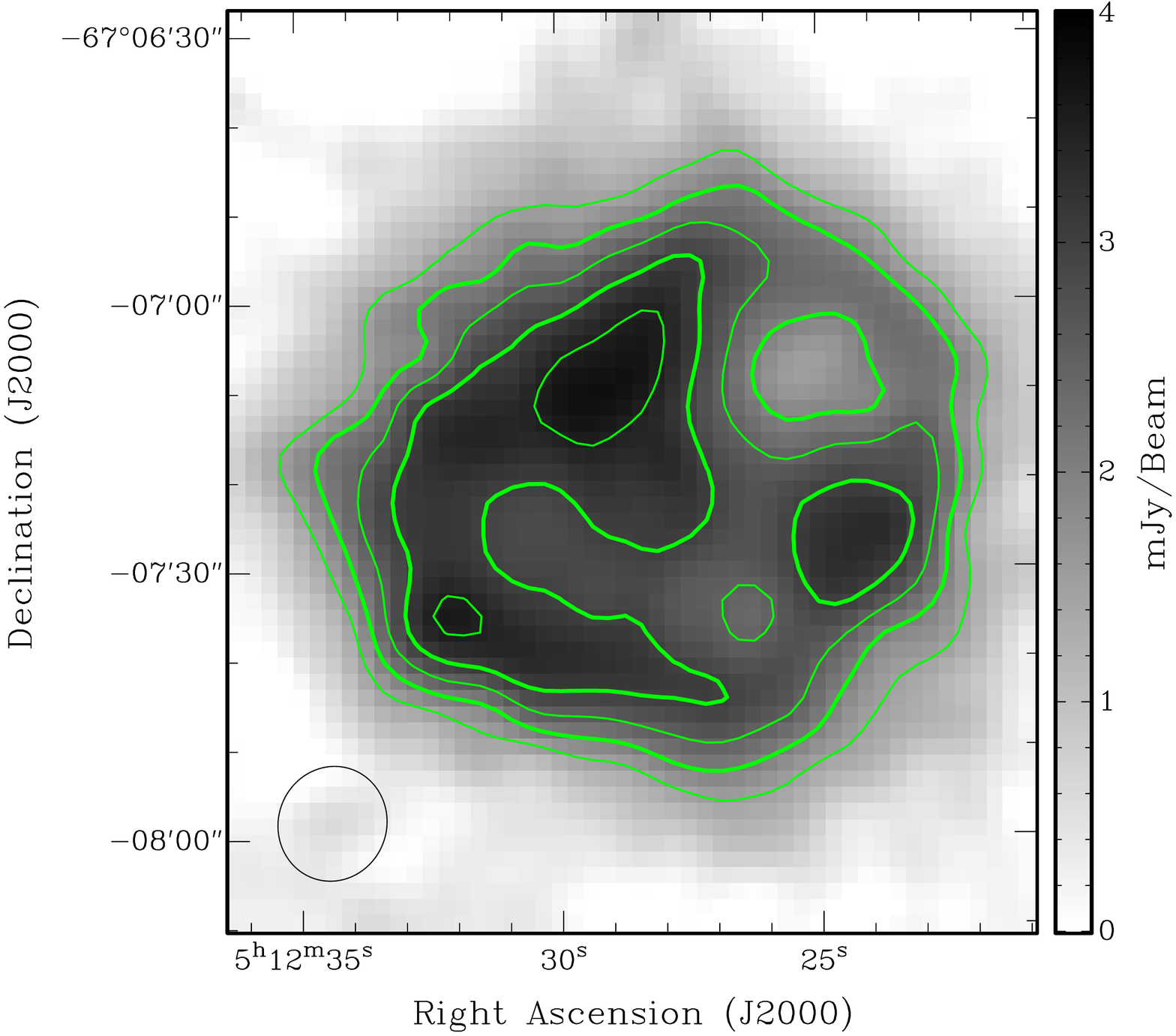}}
\caption{20~cm radio-continuum image of \SNR\ overlaid with contours at 1.5, 2.0, 2.5, 3.0, and 3.5~mJy/beam. The ellipse in the lower left corner represents the synthesised beamwidth of 13.0\arcsec$\times$\,12.1\arcsec.}\label{20cm}
\end{figure}

The relatively small size of the SNR coupled with the observing array focusing on the shorter baselines, led to the remnant remaining unresolved from these observations, and therefore, the radio morphology of the remnant cannot be readily seen. As an alternative, we use archival observations from the ATCA (project C1395) at $\lambda = 20$~cm ($\nu=1384$~MHz; bandwidth = 128~MHz). These observations were taken using ATCA arrays 1.5A and 6B on April~18~2005 and June~24~2005, respectively, for a combined integration time of 819.7~minutes. As these arrays focus on longer baselines, we were able to resolve the SNR, as seen in Fig.~\ref{20cm}. 

\begin{figure}
\resizebox{\hsize}{!}{\includegraphics[trim=1cm 5cm 0cm 5cm,angle=-90]{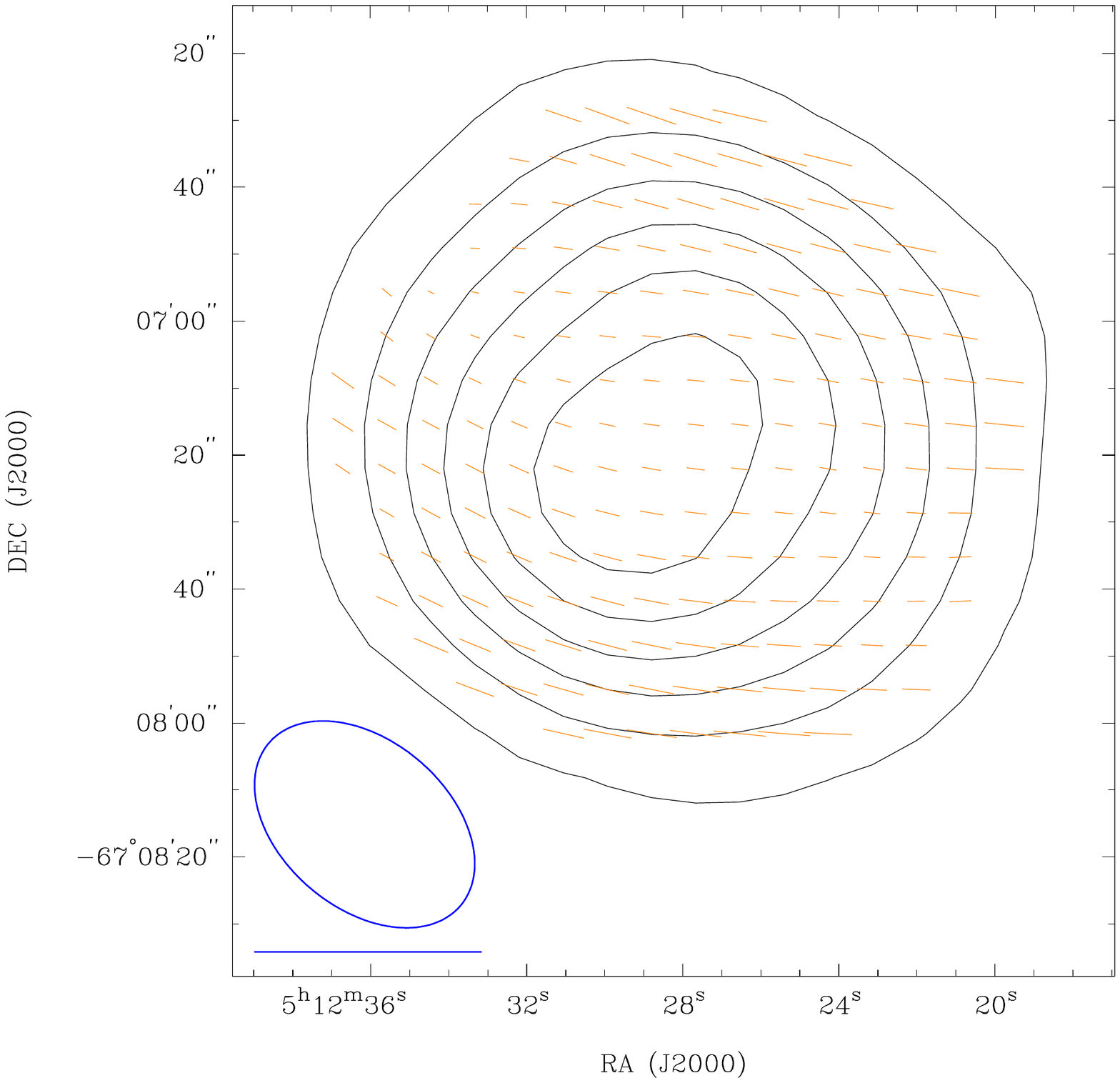}}
\resizebox{\hsize}{!}{\includegraphics[trim=1cm 5cm 0cm 5cm,angle=-90]{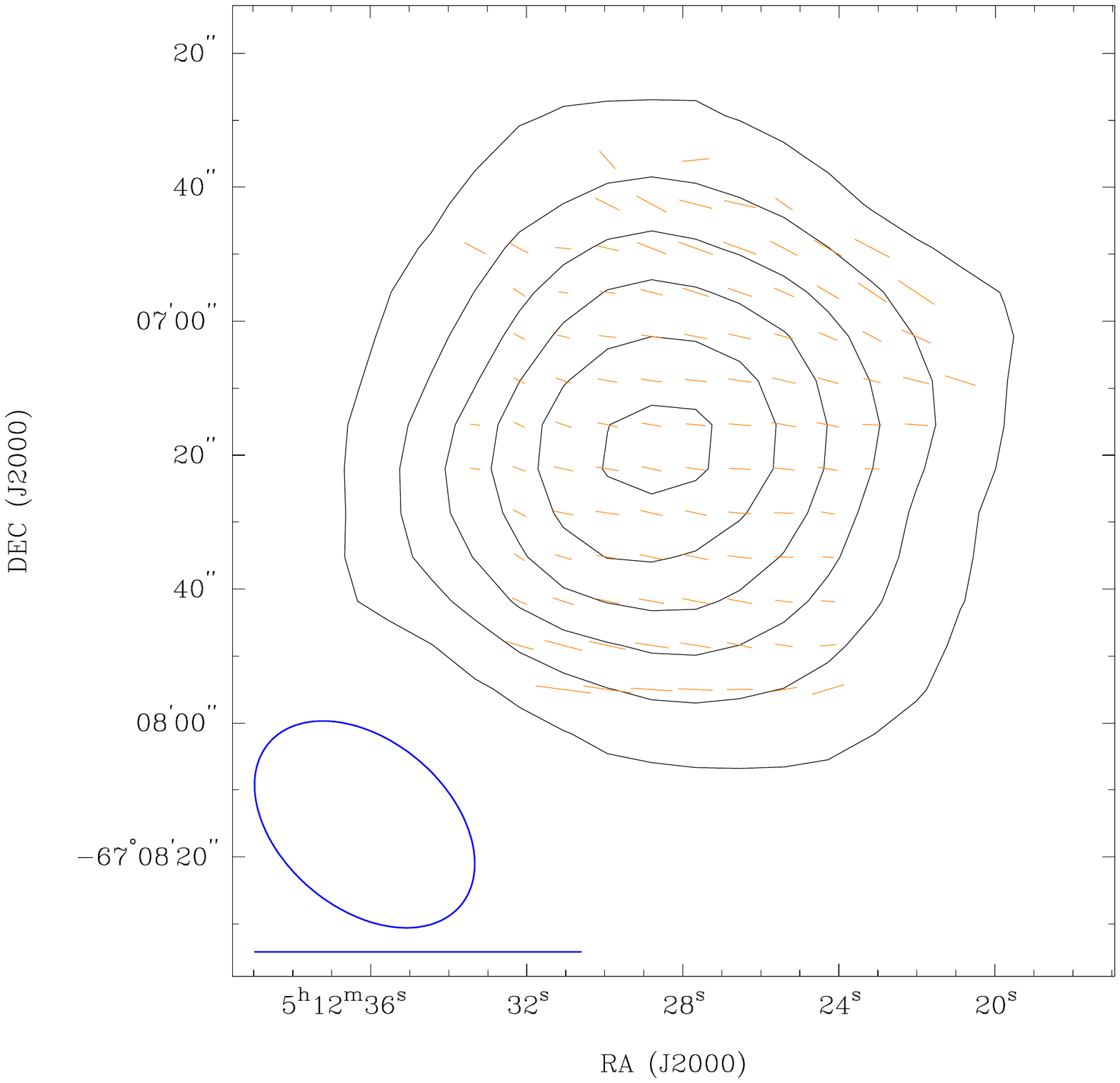}}
\caption{Polarisation of \SNR\ at 6~cm (top) and 3~cm (bottom) overlaid on an intensity image at the same wavelength. The 6~cm contours are 1 to 11~mJy in steps of 2~mJy, and the 3~cm contours are 0.9 to 9.9~mJy in steps of 1.8 mJy. The ellipse in the lower left corners represents the synthesised beamwidth of 37.5\arcsec$\times$\,25.0\arcsec, and the line directly below the ellipse represents a polarisation vector of 100 per cent.}
\label{polar}
\end{figure}

\subsubsection{CO}
To gain an understanding of the molecular environment in the region of \SNR, we used data from the Magellanic Mopra Assessment (MAGMA) survey by \citet{Wong2011}\footnote{See also \burl{http://mmwave.astro.illinois.edu/magma/}}.  MAGMA is the mapping survey of the Large and Small Magellanic Clouds in the CO(1--0) line using the 22~m Mopra Telescope of the Australia Telescope National Facility. The survey data have a spatial resolution of $\sim45\arcsec$ corresponding to 11~pc at the LMC distance.

\subsection{Infrared}
The cold environment surrounding \SNR\ can be revealed by infrared (IR) emission. To aid in the discussion of the morphology and environment of the remnant, we make use of data from the SAGE survey of the LMC \citep{Meixner2006} with the \spitzer\ \textit{Space Telescope} \citep{Werner2004}. During the SAGE survey, a $7\degr\times7\degr$ area of the LMC was observed with the Infrared Array Camera \citep[IRAC,][]{Fazio2004} in the 3.6~$\mu$m, 4.5~$\mu$m, 5.8~$\mu$m, and 8~$\mu$m bands, and with the Multiband Imaging Photometer \citep[MIPS,][]{Rieke2004} in the 24~$\mu$m, 70~$\mu$m, and 160~$\mu$m bands. The MIPS 24~$\mu$m images provide us with a picture of the stochastically, thermally, and radiatively heated dust in the region of \SNR, with spatial resolution comparable to \xmm, to give an indication of the distribution of cool material. We obtained the 24~$\mu$m MIPS mosaicked, flux-calibrated (in units of MJy~sr$^{-1}$) images processed by the SAGE team from the NASA/IPAC Infrared Science Archive\footnote{See \burl{http://irsa.ipac.caltech.edu/data/SPITZER/SAGE/}}. The pixel sizes correspond to $4.8\arcsec$ for the 24~$\mu$m band.

\section{Analysis}
The object was detected in radio, optical, and X-rays as an extended source with a size of $\sim$1\farcm7 $\times$ 1\farcm5 (see Section~\ref{mwm}). Here we describe the analysis of the SNR in these multi-wavelength datasets.

\subsection{Radio}
Linear polarisation images of \SNR\ were created at the same resolution at 6 and 3~cm using the \textit{Q} and \textit{U} Stokes parameters from the \textsc{miriad} task \texttt{IMPOL}. The mean fractional polarisation was calculated using flux density and polarisation:\\

\begin{equation}
P=\frac{\sqrt{S_{Q}^{2}+S_{U}^{2}}}{S_{I}}
\end{equation}

\noindent where $S_{Q}, S_{U}$, and $S_{I}$ are integrated intensities for the \textit{Q}, \textit{U}, and \textit{I} Stokes parameters. Mean polarisation across the remnant was found to be $12.0~(\pm1.4)$\% at 6~cm and $7.1~(\pm1.9)$\% at 3~cm. Fig.~\ref{polar} shows the polarisation for both wavelengths, where a signal-to-noise cut-off of 2$\sigma$ was used for the \textit{Q} and \textit{U} images, and a level of 6$\sigma$ for the intensity image. Values that fell below these cut-off levels are blanked in the output image. As we detected polarised emission at two frequencies, we attempted to determine the Faraday rotation through the \textsc{miriad} task \texttt{imrm}. However, as there is not any notable rotation between the vectors in the two images, the resulting errors were in the same order as the detected rotation measure.

\subsection{X-ray imaging}
\label{x-ray-imaging}
We produced images and exposure maps in various energy bands from the flare-filtered event lists for each EPIC instrument in each observation. We filtered for single and double-pixel events (\texttt{PATTERN} $\leqslant4$) from the EPIC-pn detector, with only single pixel events considered below 0.5 keV to avoid the higher detector noise contribution from the double-pixel events at these energies. All single to quadruple-pixel events (\texttt{PATTERN} $\leqslant12$) were considered for the MOS detectors. 

\par We used three energy bands suited to the analysis of the spectra of SNRs. A soft band from 0.3--0.7~keV includes strong lines from O; a medium band from 0.7--1.1~keV comprises Fe L-shell lines as well as Ne He$\alpha$ and Ly$\alpha$ lines; and a hard band (1.1--4.2~keV) which includes lines from Mg, Si, S, Ca, Ar, and non-thermal continuum if present.

\par
\begin{figure*}[!ht]
\begin{center}
\resizebox{\hsize}{!}{\includegraphics[trim= 0cm 0cm 0cm 0cm, clip=true, angle=0]{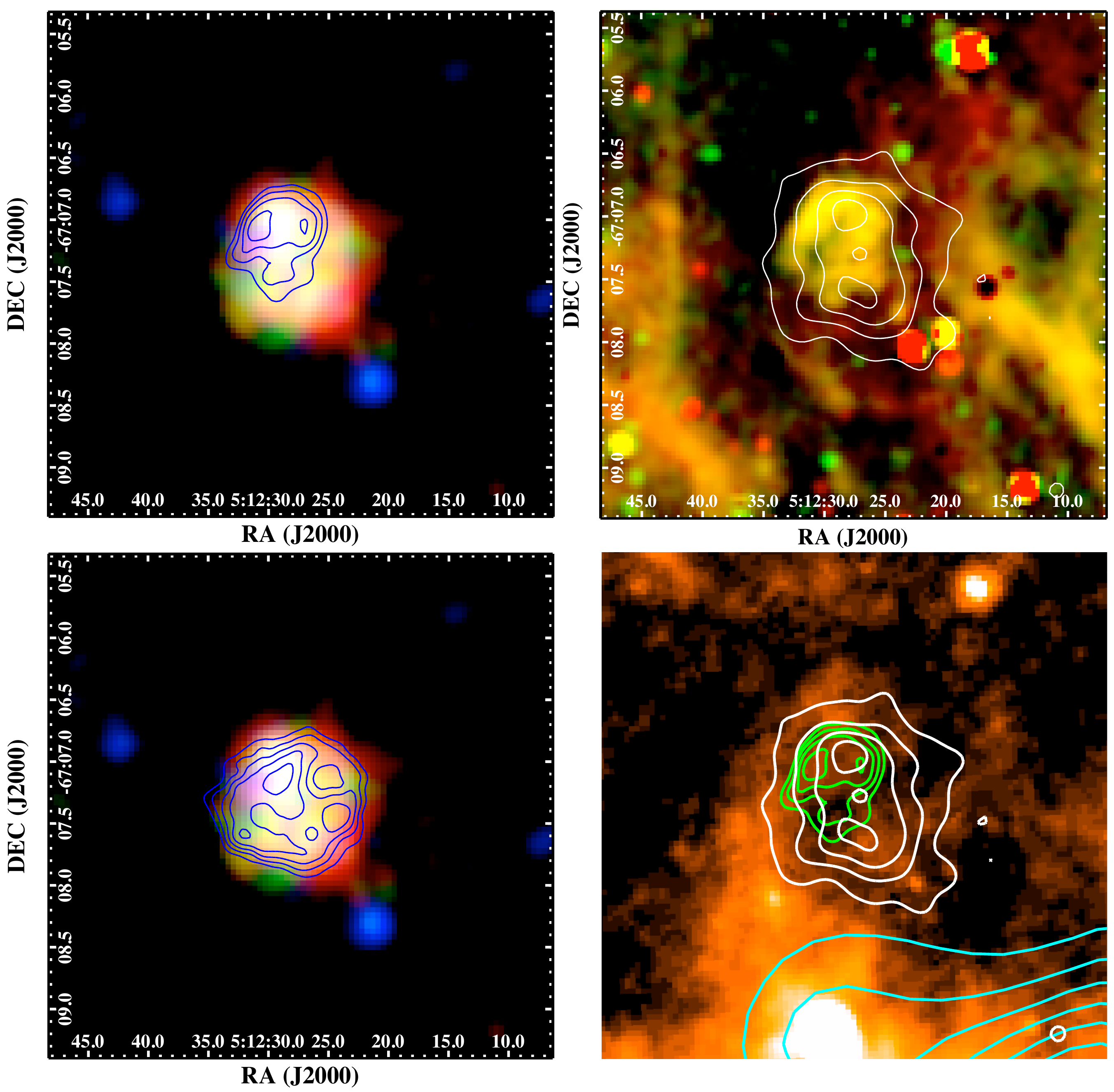}}
\caption{{\bf Top left:} \xmm\ EPIC image of \SNR\ in false colour with RGB corresponding to 0.3--0.7~keV, 0.7--1.1~keV, and 1.1--4.2~keV. The image is overlaid with \ratio\ contours with the lowest level corresponding to \sii/\ha~=~0.4, and the remaining levels at 25\%, 50\%, and 75\% of the maximum above this level. {\bf Top right:} Continuum subtracted MCELS image of \SNR\ with \ha\ in red and \sii\ shown in green overlaid with 0.3--0.7~keV contours. The lowest contour level represents 3$\sigma$ above the average background surface brightness, with the remaining levels marking 25\%, 50\%, and 75\% of the maximum above this level. {\bf Bottom left:} Same as top left but with 20~cm radio contours from Fig.~\ref{20cm} overlaid. The contours are at 1.5, 2.0, 2.5, 3.0, and 3.5~mJy/beam.  {\bf Bottom right:} \spitzer\ MIPS 24~$\mu$m image of the \SNR\ region with the X-ray contours from top right in white and \ratio\ contours from top left in green. The cyan contours represent the CO emission from the MAGMA survey and delineate the location of the molecular cloud [WHO2011]~A126. The lowest level corresponds to 1.2~K~km~s$^{-1}$ (the approximate sensitivity limit of the survey), increasing in steps of 1~K~km~s$^{-1}$. The image scale is the same as in all other panels.}
\label{snrs_im}
\end{center}
\end{figure*}

 We subtracted the detector background from the images using filter-wheel-closed data (FWC). The contribution of the detector background to each EPIC detector was estimated from the count rates in the corner of the images, which were not exposed to the sky. We then subtracted appropriately-scaled FWC data from the raw images. We merged the EPIC-pn and EPIC-MOS images from each observation into combined EPIC images and performed adaptive smoothing of each using an adaptive template determined from the combined energy band (0.3--4.2~keV) EPIC image. The sizes of Gaussian kernels were computed at each position in order to reach a signal-to-noise ratio of five, setting the minimum full width at half maximum of the kernels to $\sim5\arcsec$. In the end the smoothed images were divided by the corresponding vignetted exposure maps. Finally, we produced a three-colour image of \SNR\, which is shown in Fig. \ref{snrs_im}-top left.

\subsection{X-ray spectral analysis}
\label{spec_analysis}
For the spectral analysis, we made use of the EPIC-pn and EPIC-MOS data. However, because of a low number of total counts, the MOS data from Obs~ID~0741800201 were omitted from our analysis. We extracted source and background spectra from vignetting-weighted event lists for each EPIC instrument which include a correction for the effective area variation across the source and background. This was achieved using the SAS task \texttt{evigweight}. The SNR spectra were extracted from elliptical regions encompassing the X-ray extent of the SNR. To ensure enough counts for a solid characterisation of its components, the background spectra were extracted from a larger annulus surrounding the SNR with point sources excluded. All spectra were rebinned so that each bin contained a minimum of 30 counts to allow the use of the $\chi^{2}$ statistic during spectral fitting. The EPIC-pn and EPIC-MOS source and background spectra from each observation were fitted simultaneously using XSPEC \citep{Arnaud1996} version 12.8.2p with abundance tables set to those of \citet{Wilms2000}, photoelectric absorption cross-sections set to those of \citet{Bal1992}, and atomic data from ATOMDB~3.0.1\footnote{\burl{http://www.atomdb.org/index.php}} with the latest equilibrium and non-equilibrium data.

\subsubsection{X-ray background}
\label{x-ray_background}
Detailed descriptions of the X-ray background constituents and spectral modelling can be found in \citet{Bozzetto2014} and \citet{Maggi2014}. Here we briefly summarise the treatment of the X-ray background in the case of \SNR.

The X-ray background consists of the astrophysical X-ray background (AXB) and particle induced background.
The AXB typically comprises four or fewer components \citep{Snowden2008,Kuntz2010}, namely the unabsorbed thermal emission from the Local Hot Bubble, absorbed cool and hot thermal emission from the Galactic halo, and an absorbed power law representing unresolved background active galactic nuclei (AGN). The spectral properties of the background AGN component were fixed to the well known values of $\Gamma \sim 1.46$ and a normalisation equivalent to 10.5 photons~cm$^{-2}$~s$^{-1}$~sr$^{-1}$ at 1~keV \citep{Chen1997}. The foreground absorbing material comprises both Galactic and LMC components. The foreground Galactic absorption component was fixed at $5.1\times10^{20}$~cm$^{-2}$ based on the \citet{Dickey1990} HI maps, determined using the HEASARC $N_{\rm{H}}$ Tool\footnote{\burl{http://heasarc.gsfc.nasa.gov/cgi-bin/Tools/w3nh/w3nh.pl}}, while the foreground LMC absorption component, with abundances set to those of the LMC, was allowed to vary in the spectral fits. 

The particle-induced background of the EPIC consists of the quiescent particle background (QPB), instrumental fluorescence lines, electronic read-out noise, and residual soft proton (SP) contamination. To determine the contribution of these components we made use of vignetting corrected FWC data. We extracted FWC spectra from the same detector regions as the observational source and background spectra. The EPIC-pn and EPIC-MOS FWC spectra were fitted with the empirical models developed by \citet{SturmPhD} and \citet{Maggi2015}, respectively. Since these spectral components are not subject to the instrumental response, we used a diagonal response in XSPEC. The resulting best-fit model was included and frozen in the fits to the observational spectra, with only the widths and normalisations of the fluorescence lines allowed to vary. We also included a multiplicative constant to normalise the continuum to the observational spectra using the high energy tail ($E>5$~keV) where the QPB component dominates. The residual SP contamination was fitted by a power law not convolved with the instrumental response \citep{Kuntz2008}, which was only required in the spectra of Obs~ID~0741800301.

\subsubsection{Source emission}
Depending on the age and ambient ISM density of an SNR, the X-ray emission is expected to be dominated by shock-heated ISM swept-up by the blast wave and/or reverse shock-heated ejecta. The swept-up ISM component should have abundances consistent with the LMC, whereas the ejecta component will be characterised by strong emission lines from its metal content. A cursory scan of the extracted spectra showed emission lines below 1~keV, however, these did not appear to be significantly enhanced and do not suggest a shocked-ejecta component. The spectrum above 1~keV was notably featureless with no evidence for emission lines.

\begin{figure*}[!t]
\begin{center}
\resizebox{\hsize}{!}{\includegraphics[trim= 0cm 0cm 0cm 0cm, clip=true, angle=0]{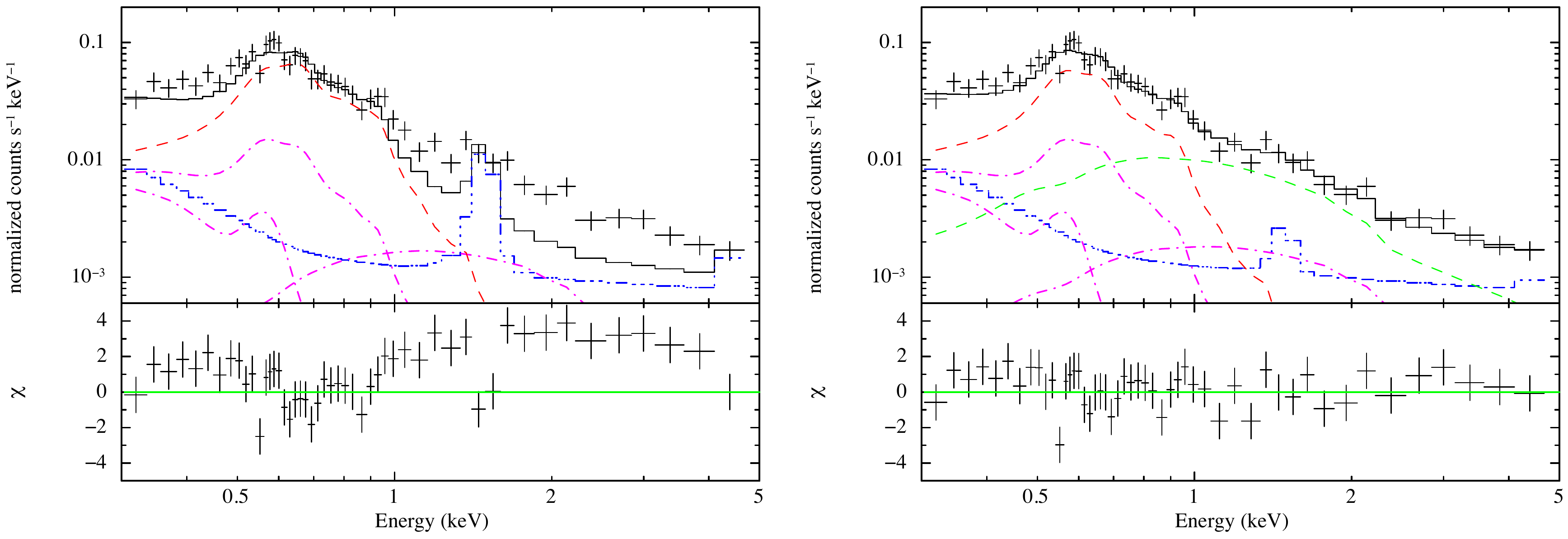}}
\caption{\xmm\ EPIC-pn spectrum of \SNR\ from Obs.~ID~0741800201, shown alone for clarity. In each panel the  best-fit model is given by the solid black line, the purple dash-dot lines mark the AXB components, the blue dash-dot-dot-dot line shows the combined contributions of the QPB, residual SPs, instrumental fluorescence lines, and electronic noise. {\bf Left:}  Best-fit \vapec\ model shown by the dashed red line. {\bf Right:}  Best-fit \vapec+\pow\ model shown by the dashed red line and dashed green line, respectively. The model fit parameters are given in Table~\ref{0533_tab}. Note that the instrumental fluorescence line normalisations are free in each fit and adjust themselves according to the residuals at their respective energies. 
}
\label{0533-spec}
\end{center}
\end{figure*}

As a first attempt to model the X-ray emission from \SNR, we included a single thermal plasma model of LMC abundance absorbed by foreground Galactic and LMC material on top of the AXB and particle-induced background. We initially assumed that the plasma was in collisional ionisation equilibrium (CIE), and therefore applied the \vapec\ \citep{Smith2001} model.  The abundance of the LMC absorption component was fixed to the LMC values \citep[$0.5~\rm{Z}_\sun$,][]{Russell1992}. The best-fit \vapec\ model (see Table~\ref{0533_tab} for fit results), with $kT \sim 0.2$~keV, adequately accounts for the emission below $\sim1$~keV but cannot explain the harder emission above 1~keV (see Fig.~\ref{0533-spec}-left). This is reflected in the fit-statistic with reduced $\chi^{2}$ ($\chi^{2}_{\nu}$) = 1.31. In initial test fits we also allowed the O and Fe abundance parameters in the \vapec\ component to vary to search for signatures of ejecta emission, with O and Fe being the dominant constituents of CC and Type~Ia explosive nucleosynthesis products, respectively. However, the resulting confidence ranges on these parameters were consistent with the LMC values, and we therefore kept the fixed LMC abundances. 

We relaxed our assumption of a CIE plasma and made use of a non-equilibrium ionisation (NEI) model in XSPEC appropriate for SNRs, namely the plane-parallel \texttt{vpshock} model \citep{Borkowski2001}. The \texttt{vpshock} model features a linear distribution of ionisation ages behind the shock which is more realistic than single ionisation age models such as \texttt{vnei}. While this model resulted in a substantially improved fit-statistic ($\chi^{2}_{\nu} = 1.11$, the best-fit model suggested a very high plasma temperature ($kT>4$~keV) and very low ionisation parameter ($\tau_{u} < 10^{10}$~s~cm$^{-3}$). Such spectral parameters are necessary to account for the hard tail above 1~keV. Allowing abundance parameters to vary to account for possible ejecta signatures did not improve the fit. Given the very high plasma temperature and very low ionisation parameter, we were reluctant to take the fit results as the true physical description of the SNR emission, though it was clear that some component must be introduced to account for the harder emission above 1~keV.

Therefore, we applied a simple two component fit with a thermal plasma and an additional power law (\pow\ in XSPEC). In trial fits we applied both the \vapec\ and \vpshock\ model to represent the softer emission component. However, since the plasma temperature and ionisation parameter (though poorly constrained) are consistent with a plasma in CIE, we proceed using the \vapec\ model to account for the soft component. The \vapec+\pow\ model resulted in an improved fit over both single thermal plasma models with $\chi^{2}_{\nu} = 1.04$. The \vapec\ model parameters are largely consistent with the single \vapec\ fit ($kT \sim 0.2$~keV), and the harder emission is accounted for by the power law with $\Gamma = 2.07~(1.81-2.26)$, see Table~\ref{0533_tab} and Fig.~\ref{0533-spec}-right. We note here that we also tried a two-temperature fit employing the \vpshock\ model for the harder component. While this yielded a good fit ( $\chi^{2}_{\nu} = 1.10$) the best-fit temperature and ionisation parameter of the \vpshock\ component were poorly constrained with the best-fit driven to high temperature and low ionisation parameter to handle the hard, featureless emission. However, we note that the hard emission may be modelled with either a power law or a thermal plasma.

\begin{table}[t]
\caption{Spectral fit results \SNR. See text for description of the models.}
\begin{center}
\label{0533_tab}
\begin{tabular}{llr}
\hline
Component & Parameter & Value\\
\hline
\hline
\multicolumn{3}{c}{\vapec}\\
\hline
\multicolumn{3}{c}{ }\\
\phabs & $N_{\rm{H,Gal}}$ ($10^{22}$ cm$^{-2}$) &   0.05\tablefootmark{a}  \\
\vphabs & $N_{\rm{H,LMC}}$ ($10^{22}$ cm$^{-2}$) &   0.39 (0.32--0.48)\tablefootmark{b}  \\
\multicolumn{3}{c}{ }\\
\vapec & $kT$ & 0.19 (0.18--0.20)\tablefootmark{b}  \\
 & $Z/\rm{Z}_{\sun}$ & 0.50 (fixed) \\
 & $EM$ ($10^{58}$ cm$^{-3}$) & 1.0 (0.7--1.6) \\
\multicolumn{3}{c}{ }\\
 & $F_{X}$\tablefootmark{c} ($10^{-14}$ erg~s$^{-1}$~cm$^{-2}$) & 2.9 \\
 & $L_{X}$\tablefootmark{d} ($10^{34}$ erg~s$^{-1}$) & 6.4 \\
\multicolumn{3}{c}{ }\\
Fit statistic & $\chi^{2}_{\nu}$ & 1.31 (736 d.o.f.) \\
\multicolumn{3}{c}{ }\\
\hline
\multicolumn{3}{c}{\vapec\ + \texttt{powerlaw}}\\
\hline
\multicolumn{3}{c}{ }\\
\phabs & $N_{\rm{H,Gal}}$ ($10^{22}$ cm$^{-2}$) &   0.05\tablefootmark{a}  \\
\vphabs & $N_{\rm{H,LMC}}$ ($10^{22}$ cm$^{-2}$) &   0.30 (0.23--0.41)\tablefootmark{b}  \\
\multicolumn{3}{c}{ }\\
\vapec & $kT$ & 0.18 (0.15--0.19)\tablefootmark{b}  \\
 & $Z/\rm{Z}_{\sun}$ & 0.50 (fixed) \\
 & $EM$ ($10^{58}$ cm$^{-3}$) & 0.7 (0.5--1.5) \\
\multicolumn{3}{c}{ }\\
\texttt{powerlaw} &  $\Gamma$ & 2.07 (1.81--2.26) \\ 
 & $norm$ ($10^{-5}$) & 1.25 (1.01--1.59) \\
\multicolumn{3}{c}{ }\\
 & $F_{X, total}$\tablefootmark{c} ($10^{-14}$ erg~s$^{-1}$~cm$^{-2}$) & 6.8 \\
 & $L_{X,\vapec}$\tablefootmark{d} ($10^{34}$ erg~s$^{-1}$) & 3.9 \\
 & $L_{X,\texttt{powerlaw}}$\tablefootmark{d} ($10^{34}$ erg~s$^{-1}$) & 2.0 \\
 \multicolumn{3}{c}{ }\\
 Fit statistic & $\chi^{2}_{\nu}$ & 1.04 (734 d.o.f.) \\
\multicolumn{3}{c}{ }\\
 \hline
\end{tabular}
\tablefoot{The numbers in parentheses are the 90\% confidence intervals.
\tablefoottext{a}{Fixed to the Galactic column density from the \citet{Dickey1990} HI maps.}
\tablefoottext{b}{Absorption and thermal component abundances fixed to those of the LMC.}
\tablefoottext{c}{0.3-10~keV absorbed X-ray flux.}
\tablefoottext{d}{0.3-10~keV de-absorbed X-ray luminosity, adopting a distance of 50~kpc to the LMC.}
}
\end{center}
\end{table}%

\section{Results}
\label{results}

\subsection{Multi-wavelength morphology}
\label{mwm}
The X-ray emission from \SNR\ is approximately circular, though slightly extended in the northeast-southwest direction, with the brightest region of emission in the northeast  (Fig.~\ref{snrs_im} top-left). Emission from the remnant is detected in all energy bands with the softest emission tracing the outer edge, and the hardest emission localised in the northeastern region and correlated with the brightest radio emission (Figs.~\ref{snrs_im} bottom-left and \ref{20cm}) and the enhanced \sii\ emission detected in the optical (Figs.~\ref{snrs_im} top-left and top-right). The MIPS~24~$\mu$m image (Fig.~\ref{snrs_im} bottom-right) reveals a possible reason for this. The remnant is projected against a denser eastern region, as evident by a large arm of cool material emanating from the [WHO2011]~A126 molecular cloud to the south and southwest, bordering the eastern front of the SNR, as noted by \citet{Reid2015}. If indeed the SNR is expanding into this cool material to the east, the higher density in the east compared to the west can explain the observed multi-wavelength morphological properties of \SNR. There is no evidence that the SNR is interacting with [WHO2011]~A126 itself.

The 20~cm radio morphology exhibits a more shell-like structure than the X-ray or optical. The radio polarisation images (Fig.~\ref{polar}), the 6~cm in particular, indicate a higher degree of polarisation in the northwest and southeast, tangentially oriented to the SNR shock front. Such a polarisation structure typically indicates evolved SNR shells \citep{Furst2004}, with the tangentially oriented magnetic fields indicative of compression of the ISM threading magnetic field. 

\begin{figure}[!t]
\begin{center}
\resizebox{\hsize}{!}{\includegraphics[trim= 0.3cm 0.8cm 2.8cm 2.2cm, clip=true, angle=0]{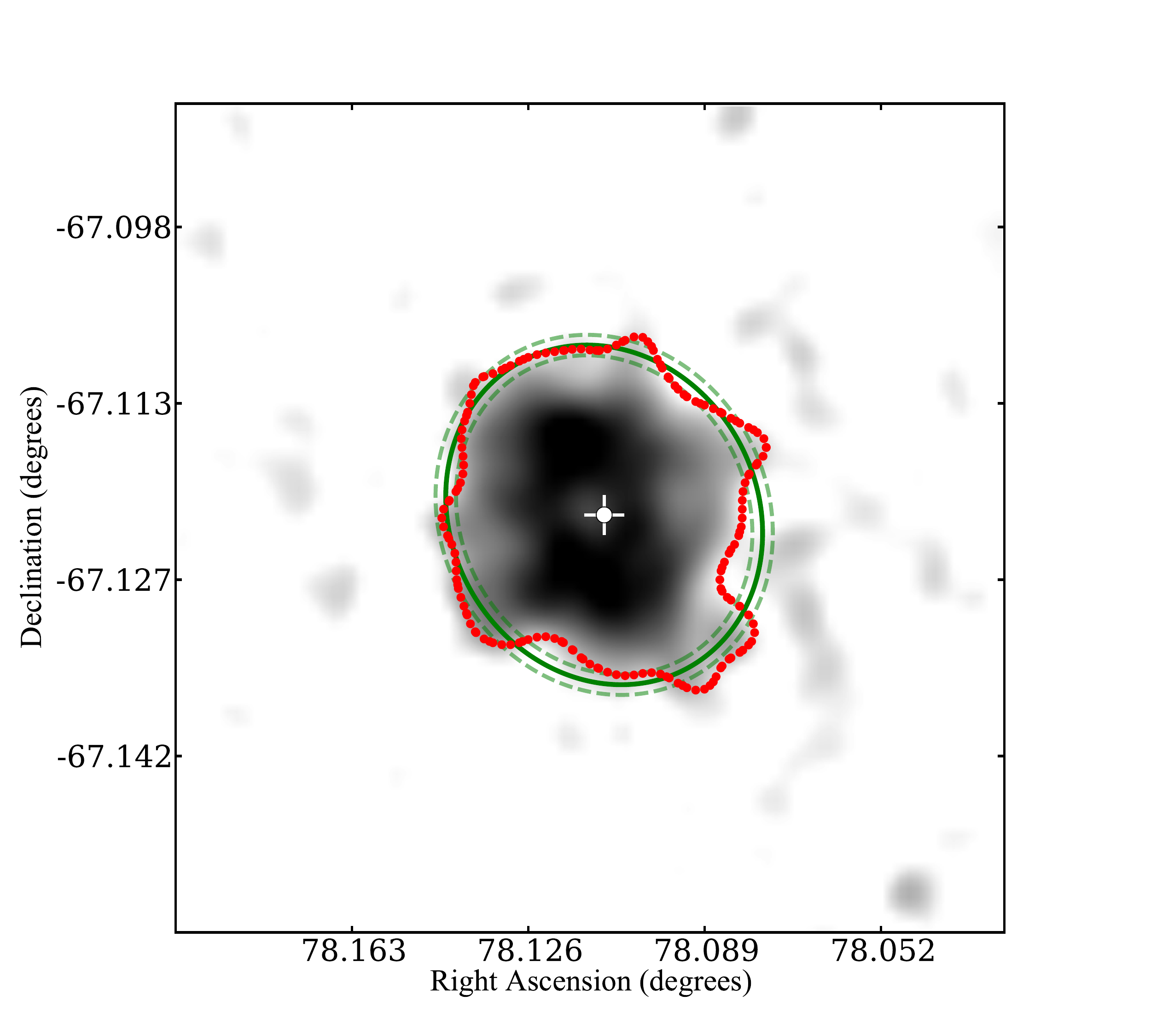}}
\caption{Combined 0.3--0.7~keV EPIC image of \SNR. The red points delineate the contour level corresponding to 3$\sigma$ above the average background surface brightness. The green solid line shows the best-fit ellipse to the contour, with the dashed lines indicating the 1$\sigma$ error on the fit. The white plus-sign marks the best-fit centre of the SNR.}
\label{size}
\end{center}
\end{figure}

To estimate the size of the X-ray remnant we determined the average background surface brightness and corresponding standard deviation ($\sigma$) in the 0.3--0.7~keV band. We defined the edge of the SNR as regions where the extended emission surface brightness rises to 3$\sigma$ above the average background and fitted an ellipse to this contour. The error on the fit was determined by quantifying the standard deviation of points on the contour from the best-fit ellipse. We determined a best-fit ellipse centred on the J2000 coordinates RA~=~05$^{\rm{h}}$12$^{\rm{m}}$26.87$^{\rm{s}}$ and Dec~=~$-67$$^{\rm{d}}$07$^{\rm{m}}$21.7$^{\rm{s}}$, of size $1\farcm72~(\pm0\farcm10)\times1\farcm51~(\pm0\farcm10)$, corresponding to $24.9~(\pm1.5)\times21.9~(\pm1.5)$~pc at the LMC distance, with the major axis rotated $\sim29^{\circ}$ east of north. The best-fit dimensions and error are shown in Fig.~\ref{size}.

\subsection{X-ray emission}
\label{xre}
The X-ray spectrum of \SNR\ is rather puzzling. There are obvious thermal features below $\sim1$~keV, which are easily fitted with a thermal plasma component in CIE. However, this leaves significant residuals above $\sim1$~keV. An NEI plasma can account for both the soft and hard parts of the spectrum, but the resulting parameters do not provide a satisfactory physical interpretation  of the X-ray emission. The addition of a second component in the form of a power law, provides the best fit to the spectrum, though again presents problems in the physical interpretation, which we now discuss.

\subsubsection{Soft emission}
We found no spectral signatures of ejecta, suggesting that the soft thermal plasma is dominated by swept-up ISM. This, and the low plasma temperature ($kT\sim0.2$~keV), are consistent with an SNR in the Sedov phase of its evolution. Using the best-fit \vapec\ parameters for both the single \vapec\ and \vapec+\pow\ models, and the Sedov solution, we estimated some properties of the remnant \citep[see][for example]{Sasaki2004,Kavanagh2015b}. However, for both the one and two component models, the derived SNR properties indicate a breakdown of the Sedov model with the determined upper limit on initial explosion energy an order of magnitude less than the canonical $10^{51}$~erg. Therefore, the soft thermal component of the \SNR\ spectrum is not consistent with an SNR in the Sedov phase.

If the remnant is not in the Sedov phase, then, given its relatively small size of $\sim25\times22$~pc, we might expect to see at least some ejecta contribution. One possible explanation that can explain both the soft thermal emission of LMC abundance and the lack of ejecta emission is that the blast wave may have recently encountered a shell of dense material which it is now propagating into. This would be the case if \SNR\ resulted from a CC explosion and the progenitor stellar wind has modified the ambient medium creating a large, low-density cavity surrounded by a dense shell of ISM. Such stellar wind blown bubbles can have radii of the order of 10~pc \citep[e.g.,][]{Weaver1977,Chen2013}, which is in agreement with the radius of \SNR\ of $\sim12$~pc. In such a scenario, the remnant would have initially swept-up the circumstellar medium, then expanded into the low-density wind-blown cavity, before interacting with the dense shell \citep{Dwarkadas2005}. When the blast wave encounters the dense shell, a shock is transmitted into the shell with significantly reduced velocity and a reflected shock travels back into the low-density interior which contains the ejecta. However, as shown by \citet{Dwarkadas2013} applied to the case of the Galactic remnant \object{Kes~27}, the emissivity of the shocked ejecta is much lower than the shocked shell material, given the dramatic difference in densities. Therefore, only the shocked-shell emission would be observed which, given reduced transmitted shock velocity and the shell composition, would exhibit a soft thermal spectrum with ISM abundances. In Section~\ref{pt}, we refer to secondary evidence in the literature that \SNR\ does indeed result from a CC explosion and, therefore, the early evolution of the SNR into a wind-blown cavity is possible.

\subsubsection{Hard emission}
The origin of the hard emission presents a significant problem in its physical interpretation. We initially considered that the hard component may be due to a point source embedded in the extended emission of the SNR, most likely a background active galactic nucleus (AGN). During our analysis we performed a source detection using the SAS task \texttt{edetect\_chain} \citep[see][for a full description of the detection method]{Kavanagh2015b}. However, the resulting sources located within the extent of SNR were identified as false detections due to the extended emission. We also searched in AGN catalogues for sources located within the extent of \SNR\ but found nothing. A further test of a possible AGN origin lies in the X-ray spectrum. If the hard component is the result of a background AGN, we should expect that the hard component is more absorbed than the soft thermal component. Therefore, we performed spectral fits allowing for the additional absorption of the hard component. However, the hydrogen column density tended to zero, with an upper limit of $\sim1\times10^{21}$~cm$^{-2}$, indicating no additional absorption of the hard emission. For these reasons, we conclude that the hard X-ray emission is associated with \SNR\ and not due to a point source. 

The hard emission is well fitted with either a simple power law or a high-$kT$, low-$\tau$ thermal plasma model of LMC abundance. Unfortunately, we are hampered in statistically distinguishing between these models because of the low count statistics at energies above $\sim2$~keV. Therefore, we now discuss possible non-thermal and thermal origins of the emission.\\

\begin{figure}[!t]
\begin{center}
\resizebox{\hsize}{!}{\includegraphics[trim= 0cm 0cm 0cm 0cm, clip=true, angle=0]{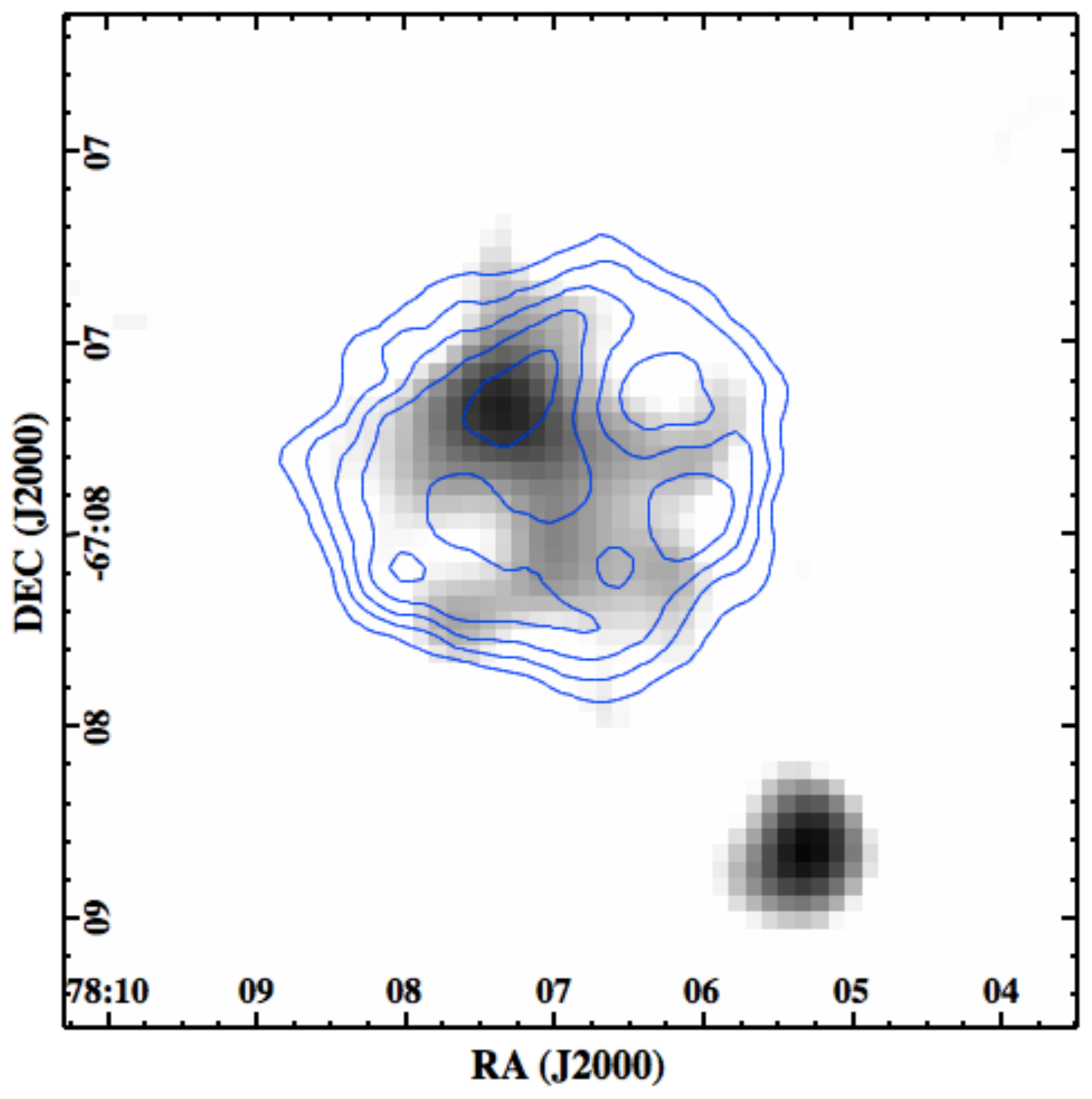}}
\caption{\xmm\ EPIC image of \SNR\ in the 2.0--4.5~keV range.  The image was produced following the same procedure outlined in Section~\ref{x-ray-imaging} and is overlaid with 20~cm radio contours from Fig.~\ref{snrs_im} - bottom left.}
\label{hard}
\end{center}
\end{figure}

\noindent \textsf{Non-thermal}\\
Morphologically, the SNR emission above $\sim2$~keV is enhanced towards the northeast of the remnant, with the brightest hard X-ray emission correlated with the brightest region of radio emission (see Fig.~\ref{hard}). In addition, this approximately correlates with the region where the polarisation fraction of the magnetic field is low and the magnetic field is oriented quasi-parallel to the SNR shock front (see Fig.~\ref{polar}), conditions favourable for particle acceleration and magnetic turbulence generation \citep{Reynoso2013}. This suggests that the hard X-ray emission could be due to synchrotron emission from very high energy electrons. However, shock speeds of the order of $10^{3}$~km~s$^{-1}$ are still required for diffusive shock acceleration and X-ray synchrotron emission \citep[see, e.g.,][]{Vink2012}. The temperature of the soft thermal X-rays is not compatible with such shock speeds. The reflected shock velocity may be high enough, especially relative to the expanding ejecta which it is moving into, though it is unclear if the synchrotron emissivity could explain the observed luminosity.
 
\begin{figure}[!t]
\begin{center}
\resizebox{\hsize}{!}{\includegraphics[trim= 0cm 0cm 0cm 0cm, clip=true, angle=0]{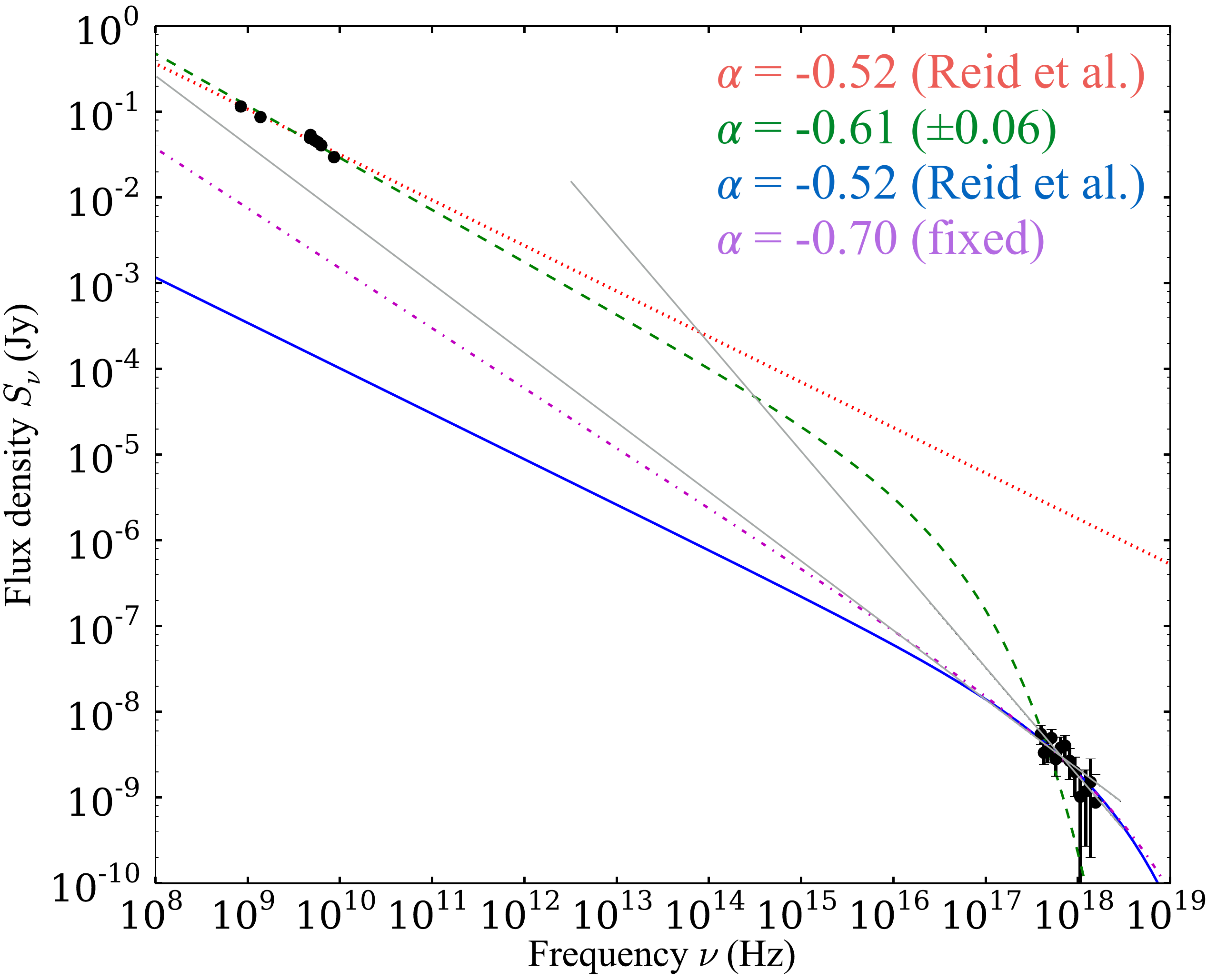}}
\caption{SED of \SNR. The data points at the lower frequencies are the radio flux densities reported in \citet{Reid2015}. The data at the higher frequencies correspond to the background-subtracted EPIC-pn spectrum from Obs~ID~0741800201 in the 2--7~keV range, which excludes the contribution of the soft X-ray component. The red dotted line shows the power-law fit to the radio data, the green dashed line shows the exponentially cut-off power law fit to the entire SED, the magenta dash-dot line shows the exponentially cut-off power law fit to the X-ray data with $\alpha$ fixed to $-0.7$, and the blue solid line shows the exponentially cut-off power law fit to the X-ray data with $\alpha$ fixed to --0.52 determined by \citet{Reid2015}. The grey lines indicate the upper and lower limit of the photon index in the X-ray power law fit (see Table~\ref{0533_tab}).}
\label{sed}
\end{center}
\end{figure}

To assess if the radio and X-ray data are compatible with the emission from a single electron population, we created a spectral energy distribution (SED). Integrated flux densities in the radio regime were reported in \citet[][their Table~4]{Reid2015}. We adopt these values for the radio points on the SED. For the X-ray points, we used the EPIC-pn spectrum from Obs.~ID~0741800201, the EPIC spectrum with the most counts. We then subtracted the background spectrum to ensure as much as possible that only X-rays due to \SNR\ were present. We confined our analysis to the $2-7$~keV energy range to isolate the portion of the spectrum where the hard component is dominant. The radio and X-ray data points are shown in Fig.~\ref{sed}.

We initially fitted the SED with a straight power law but it was immediately clear the X-ray fluxes were much lower than suggested by the radio data, which is expected if the electrons are close to the maximum energy of the relativistic electron distribution. We therefore applied a cut-off electron distribution of the form $N_{e}(E) = KE^{-s}e^{-E/E_{max}}$, where $E$ is the electron energy, $E_{max}$  the cut-off energy, $s$ is the spectral index of the electron distribution, and $K$  a constant. The spectral index of the electron distribution is related to the spectral index of the photon distribution $\alpha$ as $\alpha = (s-1)/2$. We make the simplifying assumption that each electron emits all its energy at its characteristic frequency (the $\delta$-function approximation), and thus the resulting photon spectrum cuts off as $e^{-(\nu/\nu_{max})^{1/2}}$ \citep{Reynolds1998}. This cut-off function provides a very poor fit to the SED of \SNR\ with $\alpha = -0.61~(\pm0.06)$, and $\nu_{max}=1.8~(\pm0.8) \times 10^{16}$~Hz (shown by the green dashed line in Fig.~\ref{sed}). Indeed, this fit is incompatible with the photon index confidence intervals determined in the fits to the hard X-ray emission (see Table~\ref{0533_tab}). These confidence intervals are shown by the grey lines in Fig.~\ref{sed}. Therefore, we concluded that the combined radio and hard X-ray SED is not derived from the same electron population. This might be the case of the hard X-ray emission originates from specific regions of the SNR, as already indicated by its morphology. In the case of the SNR interaction with a dense shell, as suggested by the soft thermal X-rays, the integrated radio emission is made up of synchrotron emission from electrons behind the transmitted shock and any high velocity shocks, if present. The hard X-rays, on the other hand, are emitted near high velocity shocks only. We attempted to quantify the expected radio flux densities from electrons behind the high velocity shocks as follows: we adopted the radio spectral index of \citet{Reid2015} ($\alpha=-0.52$), which is consistent with the typical value for SNRs (shown by the red dotted line in Fig.~\ref{sed}); we then fixed the value of $\alpha$ in the cut-off function to --0.52 and applied this model to the X-ray data only, resulting in the blue curve shown in Fig.~\ref{sed}. Comparing the blue curve to the radio data points indicates that the expected radio flux densities for $\alpha=-0.52$ are more than two orders of magnitude lower than observed. However, we expect this to be a lower limit since young remnants with very fast shocks have radio spectral indices of $\alpha\sim-0.7$. Fixing $\alpha$ in the cut-off function to this value results in the magenta dash-dot line in Fig.~\ref{sed}, and the expected radio flux densities are now about one order of magnitude below the observed value, which is not an unreasonably low or high estimate. Unfortunately, without deeper and higher spatial resolution radio and X-ray observations, it is impossible to create SEDs for individual regions of the SNR and, therefore, to confirm if the hard X-rays result from synchrotron emission.\\

\noindent \textsf{Thermal}\\
The reflected shock may also be employed in a thermal origin for the hard X-rays. While the transmitted shock is slowed significantly as it moves into the dense shell, the reflected shock travels back through the still expanding ejecta, further heating the already shocked gas. Such a reflected shock-heating scenario has been used to explain the observed morphology at shock-cloud interaction regions in the \object{Cygnus Loop} \citep{Levenson2002}.

Another, though somewhat less likely, thermal explanation for the hard emission is the presence of a background galaxy cluster projected against the SNR. The hot intracluster medium of these objects can be described by CIE models with $kT$ values up to $\sim10$~keV \citep[for a review, see][]{Bohringer2010}. At such high temperatures, the spectrum is effectively featureless, which is especially true for low metallicity plasmas. Trial fits with a CIE plasma model in place of the power-law result in acceptable fits of equivalent goodness. However, the required chance alignment of a background galaxy cluster, located at the position of \SNR\ with a compatible morphology, makes this explanation for the hard component is extremely tenuous. In addition, the correlation of the synchrotron radio emission with Bremsstrahlung dominated galaxy cluster X-ray emission makes no physical sense. \\

We have briefly discussed some possible origins for the hard X-ray emission from \SNR\ here but we cannot offer a definitive explanation. Deep high spatial resolution observations with \chandra\ would allow the morphology of the hard emission region to be properly traced, helping to better identify the origin of the hard X-rays.

\subsection{Progenitor type}
\label{pt}
The apparent association of \SNR\ with the DEM~L97 HII region \citep{Davies1976} and the proximity to the [WHO2011] A126 molecular cloud \citep{Wong2011} would qualitatively favour a CC origin for the SNR. Unfortunately, given the lack of ejecta signatures in the X-ray spectrum, it is not possible to suggest a supernova remnant type directly.

A secondary indicator for the SN type is the local stellar population and star formation history (SFH). For example, a local stellar population harbouring many massive early-type stars and a corresponding recent burst of star formation would indicate that the SNR is most likely the result of a core-collapse explosion. Similarly, the absence of such a high mass stellar population and recent star formation burst would point to a Type~Ia origin. While this typing method is not infallible \citep[see][for a detailed description of the method and caveats]{Maggi2015} it still offers a good indication of the likely SN type and has been used in the past to gauge the explosion scenarios for a number of evolved LMC remnants \citep[e.g.,][]{Kavanagh2013,Maggi2014,Bozzetto2014}.

\citet{Maggi2015} performed a stellar population and SFH analysis of all confirmed SNRs in the LMC using photometric data from \citet{Zaritsky2004} and the spatially resolved SFH maps of \citet{Harris2009}. They identified 25 massive OB stars located within 100~pc of \SNR, inferring an expected CC/Type~Ia ratio for the region of $5.56^{+1.2}_{-2.66}$ (their Table~C.1). These factors indicate that \SNR\ most likely resulted from a CC event.

\section{Summary}
\label{sum}
We presented a study of \SNR\ in the LMC and supplemented this with optical, radio, and IR data to gain an understanding of the physical properties of the SNR. The main findings of our analysis can be summarised as follows:

\begin{itemize}
\item We showed that the \rosat\ source [HP99~483]  is the X-ray counterpart of the supernova remnant \SNR\ identified by \citet{Reid2015}. However, we determined that the remnant is larger than determined by \citet{Reid2015} from the optical emission. We suggested that the optical emission is confined to a region where the remnant appears to be expanding into a denser ambient medium. 

\item The 20~cm radio morphology exhibits a shell-like structure. Radio polarisation images at 3~cm and 6~cm in particular, indicate a higher degree of polarisation in the northwest and southeast, tangentially oriented to the SNR shock front, indicative of an SNR where the expanding shell has compressed the magnetic field threading the ambient ISM.

\item The X-ray spectrum of \SNR\ is unusual in that it exhibits soft emission lines which are consistent with a soft thermal plasma in collisional ionisation equilibrium, but also shows featureless hard emission. We suggested that the soft thermal X-rays can be explained by a remnant initially evolving into the wind-blown cavity of a progenitor massive star and is now interacting with the surrounding dense shell. We discuss some possible thermal and non-thermal emission mechanisms for the hard X-ray component, though the true origin of these X-rays remains unclear.  We found no spectral signatures of shocked ejecta.

\item The location of \SNR\ in the DEM~L97 HII region and at the edge of the [WHO2011] A126 molecular cloud point to a core-collapse origin to the remnant. Due to the lack of ejecta signatures, we cannot directly determine the supernova type from the X-ray spectrum. Indirect evidence for the explosion mechanism is found in the study of the local stellar population and star formation history study of \citet{Maggi2015}, which, because of the large number of nearby OB stars and recent star formation burst, suggest a core-collapse origin. This is consistent with the SNR evolution into a the wind-blown cavity suggested above.

\end{itemize}

\begin{acknowledgements} We thank the anonymous referee for reviewing our manuscript. P.J.K. acknowledges support from the Bundesministerium f\"{u}r Wirtschaft und Technologie/Deutsches Zentrum f\"{u}r Luft- und Raumfahrt (BMWi/DLR) through grant FKZ 50 OR 1309. M.S. acknowledges support by the Deutsche Forschungsgemeinschaft through the Emmy Noether Research Grant SA2131/1-1. P.\,M. acknowledges support by the Centre National d'\'Etudes Spatiales (CNES). Cerro Tololo Inter-American Observatory (CTIO) is operated by the Association of Universities for Research in Astronomy Inc. (AURA), under a cooperative agreement with the National Science Foundation (NSF) as part of the National Optical Astronomy Observatories (NOAO). We gratefully acknowledge the support of CTIO and all the assistance that was provided for upgrading the Curtis Schmidt telescope. The MCELS project has been supported in part by NSF grants AST-9540747 and AST-0307613, and through the generous support of the Dean B. McLaughlin Fund at the University of Michigan, a bequest from the family of Dr. Dean B. McLaughlin in memory of his lasting impact on Astronomy. We used the karma software package developed by the ATNF. The Australia Telescope Compact Array is part of the Australia Telescope, which is funded by the Commonwealth of Australia for operation as a National Facility managed by CSIRO. \end{acknowledgements} 

\bibliographystyle{aa}

\end{document}